\documentclass[superscriptaddress,secnumarabic,
amssymb,amsmath,nobibnotes,aps,prd,showkeys,showpacs,nofootinbib]{revtex4}%
\usepackage{graphicx}
\usepackage{epsf}
\usepackage{epsfig}
\usepackage{scrtime}
\usepackage{multirow}
\usepackage[multiple]{footmisc}
\usepackage{amsfonts, graphics}
\usepackage{bm}
\usepackage[utf8x]{inputenc}
\usepackage{amsmath}
\usepackage{amsfonts}
\usepackage{amssymb}
\usepackage{color}
\usepackage{epstopdf}
\usepackage{natbib}%
\providecommand{\U}[1]{\protect\rule{.1in}{.1in}}
\newcommand{\be}{\begin{equation}}
\newcommand{\ee}{\end{equation}}

\newcommand{\mincir}{\raise
-3.truept\hbox{\rlap{\hbox{$\sim$}}\raise4.truept\hbox{$<$}\ }}
\newcommand{\magcir}{\raise
-3.truept\hbox{\rlap{\hbox{$\sim$}}\raise4.truept\hbox{$>$}\ }}

\begin{document}
\title{Revisiting the Evolving Lorentzian Wormhole: A General Perspective}
\author{Subhra Bhattacharya}
\email{subhra.maths@presiuniv.ac.in}
\affiliation{Department of Mathematics, Presidency University, Kolkata-700073, India}
\author{Tanwi Bandyopadhyay}
\email{Tanwi.Bandyopadhyay@aiim.ac.in }
\affiliation{ Adani Institute of Infrastructure Engineering, Ahmedabad 382421}

\begin{abstract}

Wormholes can be described as geometrical structures in space and time that can serve as connection between distant regions of the universe. Mathematically, general wormholes can be defined both on stationary as well as on dynamic line elements. However, general relativistic and evolving Lorentzian wormholes are less studied than their static wormhole counterpart. Accordingly, in this work we shall focus on some evolving wormhole geometries. Starting from a general class of spherically symmetric line element supporting wormhole geometries, we shall use the Einstein's field equations to develop viable astatic wormhole solutions. We will also discuss various evolving wormhole solutions together with their physical significance, properties and throat energy conditions. We claim that the method discussed in this work shall be applicable for developing wormhole solutions corresponding to any general Lorentzian wormhole metric.
\end{abstract}
%%%%%%%%%%%%%%%%%%%%%%%%%%%%%%%%%%%%%%%%%%%%%%%%%%%%%%%%%%%%%%%%%%%%%%%%%%%%%%%%%%%%%%%%%%%%%%%%%%%%%%%%%%%%%%
\maketitle
%%%%%%%%%%%%%%%%%%%%%%%%%%%%%%%%%%%%%%%%%%%%%%%%%%%%%%%%%%%%%%%%%%%%%%%%%%%%%%%%%%%%%%%%%%%%%%%%%%%%%%%%%%%%%%%%%%
%~~~\myclassification{98.80.Cq, 98.80.-k}\\\\
%%%%%%%%%%%%%%%%%%%%%%%%%%%%%%%%%%%%%%%%%%%%%%%%%%%%%%%%%%%%%%%%%%%%%%%%%%%%%%%%%%%%%%%%%%%%%%%%%%%%%%%%%%%%%%%%%%
\section{Introduction}

Wormholes are conjectured to be hypothetical structures in space and time that can be obtained as a solution of the Einstein's field equations (EFE). Presently wormhole solutions to EFE are highly popular and well studied topics in the literature, yet the process of its development and refinement to its current form took substantial period. The first discussion on wormholes was due to Flamm \cite{flamm}, back in 1916. He mentioned them as a contribution of Einstein's theory of gravity. Couple of decades later in 1935, Einstein and Rosen mentioned them as ``shortcut" or bridges between distant regions of the universe \cite{er}. They were later popularised as the Einstein-Rosen bridge. This was followed by Wheeler's work on ``Geometrodynamics" \cite{wh}, where he laid down the geometrical idea of ``doubly connected" distant regions of the universe. After another couple of years Misner and Wheeler \cite{mw} finally called such ``handles" in space time as ``wormholes." In subsequent work, charged scalar field solutions to wormhole like structures were provided by Bronnikov \cite{bronn} and by Kodama \cite{kod}. Kodama mentioned them as the ``kink" solutions. Around the same time, another class of scalar field solutions were given by Ellis \cite{elli},  which were called the ``drainhole". Soon after Clement considered wormhole like solutions in higher dimensional gravity theory \cite{cle}. The interesting common feature of all these solutions were that they were static wormhole like configurations. Time dependent wormhole solutions were studied much later in the literature.

In fact, it can be stated that after Morris and Thorne's generalization of a static spherically symmetric Lorentzian wormhole \cite{m+t}, non-static geometries were considered for the same. It was after this paper that the idea of traversable wormhole was popularised. The particularly interesting aspect of Morris and Thorne wormhole solution was the possibility of using them as rapid mode of transport between distant stellar objects. (Accomplishing such engineering feats would however require sophisticated physics and mathematics that bypasses hurdles like the cosmic censorship and causality violation principle). Nonetheless the question had intrigued and baffled scientists for years and remains unanswered till date. The traversable Lorentzian wormhole described in \cite{m+t} were on a stationary metric with imposed conditions of ``no horizon" and ``radial asymptotic flare". These ensured that the ensuing geometry would allow transport between space-times. However due to EFE, gravity effected the resultant matter stress energy tensor so that it violated the null energy condition (NEC). This provided a stimulus in developing wormhole solutions with matter satisfying some averaged energy conditions. Morris and Thorne mentioned ways to limit the usage of exotic matter (a form of matter that violates NEC). Ensuing literature provided wormhole solutions either in modified gravity theories or in normal gravity with matter that conformed to known energy conditions. (Please refer to \cite{visserb} for a review).

Introducing time dependent metric gives a dynamic wormhole configuration. One of the earliest mention of some construction efforts to such non-static configurations could be found in \cite{visser}. Here the author uses the technique of cut and paste with suitable juncture convergence criteria to develop a class of traversable wormholes of both static and dynamic variety. Along similar lines Hochberg \cite{hoch}, developed static and astatic wormhole in higher order $R^{2}$ gravity. The authors of \cite{hk} used such surgically modified dynamic wormhole solutions to address the horizon problem in the early inflationary universe. T. A. Roman \cite{rom} considered dynamic wormhole with a de-Sitter factor and analysed the effect of inflation on a submicroscopic dynamic wormhole. The paper in some scope addressed the pertinent question of inflating a submicroscopic wormhole for the creation of time machine, which was raised by the authors of \cite{mty}. The authors of \cite{mty} elaborate upon some of the relevant mathematical approaches to build a time machine out of a wormhole. Later literature showed that existence of such submicroscopic Lorentzian static wormhole might not be permitted using Ford-Roman quantum inequality principle \cite{kkn}. S. Kar in his paper  \cite{kar}, for the first time introduced a particularity in study of dynamic wormholes. The authors of \cite{kar} and \cite{ks}, developed specific non-static wormhole versions of their static Lorentzian counterpart. They showed that it might be possible to obtain such astatic space-time configuration conforming to wormhole geometry, but with throat matter satisfying energy conditions. A similar analysis was performed in \cite{kim}, where it was shown that dynamic wormholes in FRW space-time could exist with matter satisfying energy conditions. Hochberg and Visser \cite{hv} and Hayward \cite{hay} gave an altogether new perspective of dynamic wormhole throats as ``anti-trapped surfaces" by considering the local geometry at the throat. In such definitions, they argued that NEC violation would be a generic property of the throat which could not be thwarted.

Evolving wormhole geometries supported by phantom matter (matter with barotropic equation of state $\omega<-1$) was considered in \cite{cat1}. As a continuation of the work in \cite{cat1}, authors of \cite{cat2,cat3} showed that dynamic wormholes could be supported by fluid mixtures of more than one kind such that the total matter satisfied the energy conditions. In fact such wormholes would expand with expansion driven by the homogeneous and isotropic component of the matter mixture. They further their previous two work in \cite{cat4}, by showing that such wormholes would end in big-rip type of singularity \cite{bar} at spatial infinity provided the homogeneous component is phantom in nature. Another version of dynamic wormholes with throat matter satisfying NEC were considered in \cite{mae}. Here the wormhole was called ``cosmological" and connected two FRW universes, with the whole space time being trapped, unlike the previous version in \cite{hv,hay}, where the throat was trapped.

Recently dynamic wormholes have been studied from the perspective of particle creation mechanism \cite{pan} and in $f(R)$ gravity theory in \cite{sb}. Evolving wormhole solutions were obtained using the Mc Vittie solutions in \cite{kim1}. Part of the work in the this article could be considered as generalization of the work done in \cite{kar}. In this work we shall provide all possible dynamic wormhole cosmologies that can exist in Einstein gravity. The wormhole considered in \cite{kar,ks} shall appear as special case of our solution. We will also consider the energy conditions of the throat matter and their dependence on the temporal factor. 

The paper is organised into the following sections. In section 2 we shall describe a general evolving wormhole and discuss corresponding solutions. In section 3 we shall discuss the properties of the wormholes considered and their corresponding energy conditions. In section 4 we shall provide a numerical analysis and corresponding graphical representation of our analytical results with corresponding discussions. Finally section 5 will contain some concluding remarks.

\section{Evolving Wormhole Solution}

A general spherically symmetric line element that can contain wormhole like solutions is given by:
\begin{equation}
ds^{2}=-e^{2\phi(r,t)}dt^{2}+a^{2}(t)\left[\frac{dr^{2}}{1-\frac{b(r)}{r}}+r^{2}d\Omega^{2}\right]\label{metric}
\end{equation}
where $d\Omega^{2}=\sin^{2}\theta d\psi^{2}+d\theta^{2}$ and $a(t)$ is the usual scaling of the temporal coordinate. The above metric was conceived in \cite{m+t} for the first time. Their analysis provided a comparative  description of the black hole event horizon and how it can be tackled in a wormhole so that they are suitable as means of travel. Here $e^{2\phi(r,t)}$ is the red-shift function. Mathematically the red-shift function vanishes at the horizon and hence helps in its physical identification. Since traversable wormholes are preconditioned to be without event horizon \cite{m+t}, the red shift functions are assumed as finite everywhere. The function $b(r)$ is called the shape function, and indicates the nature of the wormhole throat. Mathematically the throat implies the surface of minimum radius. Hence if $r_{0}$ is the minimum value of the radial coordinate, then at the throat $b(r_{0})=r_{0}.$ Also the throat function $b(r)$ must satisfy the flare out condition such that $\frac{b(r)}{r}\rightarrow 0$ as $r\rightarrow\infty.$ Considering the components of anisotropic energy momentum tensors  as $(\rho(r,t),p_{r}(r,t),p_{t}(r,t)),$ the EFE for the above metric (\ref{metric}) is given by:
\begin{align}
\kappa\rho(r,t)=&3H^{2}e^{-2\phi(r,t)}+\frac{b'(r)}{a^{2}r^{2}}\label{erho}\\
\kappa p_{r}(r,t)=&-(2\dot{H}+3H^{2}-2H\phi_{t})e^{-2\phi(r,t)}-\frac{b(r)}{a^{2}r^{3}}+\frac{2(r-b(r))}{a^{2}r^{2}}\phi_{r}\label{epr}\\
\kappa p_{t}(r,t)=&-(2\dot{H}+3H^{2}-2H\phi_{t})e^{-2\phi(r,t)}+\frac{b(r)-rb'(r)}{2a^{2}r^{3}}+\frac{2r-b(r)-rb'(r)}{2a^{2}r^{2}}\phi_{r}+\frac{r-b(r)}{a^{2}r}(\phi_{r}^{2}+\phi_{rr})\label{ept}\\
0=&-2e^{-2\phi(r,t)}\phi_{r}\dot{a}\frac{r-b(r)}{r}\label{constraint}
\end{align}
where $H=\frac{\dot{a}}{a},$ \emph{dot} indicates differentiation with respect to $t$ and \emph{dash} indicates differentiation with respect to the $r.$ From the equation (\ref{constraint}) we see that it is satisfied under three different possibilities. The different possibilities can be: i) The metric can be static, inhomogeneous and spherically symmetric, which gives $\dot{a}=0,$ ii) $b(r)=r,$ which evidently is not viable under the metric consideration, and finally iii) $\phi_{r}=0,$ that is the metric is dynamic, inhomogeneous with the red-shift function being independent of the radial coordinate. So we claim that dynamic wormholes with red-shift function having dependence on the radial coordinate is in general not possible. Considering the motivation of the current work we shall restrict our choice to the third case only. Further this third restriction affects the components of the stress tensor such that we can now fix $p_{r}(r,t)=\alpha p_{t}(r,t),$ where $\alpha$ can be a constant or it can be a function of $r.$ The corresponding matter-stress energy tensors satisfy the conservation equations $T^{\nu}_{\mu;\nu}=0$ as given by:
\begin{align}
\frac{\partial\rho}{\partial t}+H(3\rho+p_{r}+2p_{t})=0\label{consv1}\\
\frac{\partial p_{r}}{\partial r}=\frac{2}{r}\left(p_{t}-p_{r}\right)\label{consv2}
\end{align}

For generality we start by assuming that $\alpha=\alpha(r),$($\alpha=constant$ will follow as a special case of the above), which gives:
\begin{equation}
a^{2}(2\dot{H}+3H^{2}-2H\dot{\phi})e^{-2\phi(t)}=-\frac{1}{(1-\alpha(r))}\left[\frac{b(r)}{r^{3}}\left(1+\frac{\alpha(r)}{2}\right)-\frac{b'(r)}{r^{2}}\left(\frac{\alpha(r)}{2}\right)\right]\label{me}
\end{equation}
The above equation has four unknowns: $\phi(t),b(r),a(t)$ and $\alpha(r).$ This can be considered as the master equation for solving the above unknowns. 

We observe that the right hand side of equation (\ref{me}) is a function of the radial coordinate $r,$ while the left hand side is a function of the temporal coordinate $t.$ Since the coordinates $t$ and $r$ are independent, we equate the left hand side and right hand side both, to a constant factor $\beta.$ This gives the following two equations: 
\begin{align}
a^{2}(2\dot{H}+3H^{2}-2H\dot{\phi})e^{-2\phi(t)}-\beta=0\label{frt}\\
b'(r)-\frac{b}{r}\left(1+\frac{2}{\alpha(r)}\right)-2\beta r^{2}\left(\frac{1}{\alpha(r)}-1\right)=0\label{frr}
\end{align}
These differential equations with unknowns $a(t)$ and $b(r)$ can be solved for some known values of $\alpha(r)$ and $\phi(t).$ 

\subsection{Solution for $a(t)$}

Equation (\ref{frt}) can be rewritten as follows:
\begin{equation}
a\ddot{a}-a\dot{a}\phi_{t}+\frac{1}{2}\dot{a}^{2}-2\omega^{2}e^{2\phi}=0\label{at}
\end{equation}
with $\omega^{2}=\frac{\beta}{4}.$ This is a non-linear second order differential equation in variable $a(t).$ We will solve the above equation as follows:

\subsubsection{Trivial Solution}
Here we assume that $e^{\phi(t)}=a(t).$ This reduces equation (\ref{at}) into a linear ordinary differential equation $\ddot{X}-2\omega^{2}X=0$ where $a(t)=X^{2}.$ Solving this we get:
\begin{description}
\item $\beta>0\Rightarrow a(t)=e^{\pm 2\omega t}$
\item $\beta<0\Rightarrow a(t)=e^{\pm 2i\omega t}$
\item $\beta=0\Rightarrow a(t)=\left(\frac{t}{t_{0}}\right)^{2}.$
\end{description}
By this choice of $e^{\phi(t)}$ we are actually replacing the conformal time $t$ with proper time $\tau$ by the equivalence $d\tau=a(t)dt.$ Accordingly $a(t)\equiv A(\tau)$ such that the wormhole metric now resembles a dynamic wormhole with zero-tidal force supported by anisotropic matter,
\begin{equation}
ds^{2}=-d\tau^{2}+A^{2}(\tau)\left[\frac{dr^{2}}{1-\frac{b(r)}{r}}+r^{2}d\Omega^{2}\right].
\end{equation}
It is evident that for the choice of $\beta>0$ the solution given by $a(t)=e^{2\omega t}$ resembles a de-Sitter scale factor and the corresponding solution giving a conformal time de-Sitter line element. This essentially gives an exponentially growing wormhole geometry \cite{rom}. Further the choice for $\beta<0$ gives a metric conformal to the anti de-Sitter metric. While for $\beta=0$ the solution of $a(t)=\left(\frac{t}{t_{0}}\right)^{2}$ will resemble a Milne universe.

\subsubsection{A More General Solution}
\begin{description}
\item By taking a substitution $a(t)=X^{2/3},$ equation (\ref{at}) can be reduced to the following equation: $X^{1/3}(\ddot{X}-\phi_{t}X)=3\omega^{2}e^{2\phi(t)}.$ For any general $\phi(t)$ this equation can be solved only for the case $\beta=0.$ The solution is obtained as $a(t)=a_{0}(\int e^{\phi(t)}dt)^{\frac{2}{3}}.$ This is the most general solution possible for different forms of the function $\phi(t).$ 

\item A third solution of (\ref{at}): We rewrite the equation as 
\begin{equation}
a\dot{a}\phi_{t}+2\omega^{2}e^{2\phi}=a\ddot{a}+\frac{1}{2}\dot{a}^{2}
\end{equation}
and set $a\ddot{a}+\frac{1}{2}\dot{a}^{2}=0.$ We solve the above coupled equations and get the solutions $a(t)=(c_{1}t+c_{2})^{2/3}$ and $e^{2\phi(t)}=\frac{c_{1}^{2}}{9\omega^{2}}(c_{1}t+c_{2})^{-1/3}.$ Here $c_{1}$ and $c_{2}$ are two constants of integration, such that $t\neq-\frac{c_{2}}{c_{1}}$. Setting $c_{2}=0$ and $c_{1}=1$ we get $a(t)=t^{2/3},$ which resembles the matter dominated Einstein de-Sitter universe.
\end{description}

\subsection{Solution for $b(r)$}

To solve for equation (\ref{frr}) with any general $\alpha(r),$ we rewrite $\alpha(r)=\frac{g(r)}{rg'(r)}$ corresponding to any arbitrary function $g(r)$ of $r.$ With this we obtain the solution for $b(r)$ as follows:
\begin{equation}
\frac{b(r)}{r}=-\beta r^{2}+g^{2}(r)\label{b}
\end{equation}
Depending upon the choice of the function $g(r)$ we obtain viable solutions for the wormhole throat. Substituting this result for $b(r)$ in the metric (\ref{metric}) gives:
\begin{equation}
ds^{2}=-e^{2\phi(t)}dt^{2}+a^{2}(t)\left[\frac{dr^{2}}{1-(-\beta) r^{2}-\frac{b_{n}(r)}{r}}+r^{2}d\Omega^{2}\right]\label{nmetric}
\end{equation}
where $\frac{b_{n}(r)}{r}=g^{2}(r).$ In the above line-element, the constant $-\beta$ can be interpreted as the curvature of space-time, which can have values 0, $-1$ and $+1$ for flat, open and closed universe in the sense of a FRW line element. Further, letting $g^{2}(r)\rightarrow 0$ (for $r\rightarrow\infty$) one actually obtains the FRW line element. Now considering the wormhole metric as defined above, the effective shape function that defines the wormhole throat is given as $\frac{b_{n}(r)}{r}=g^{2}(r).$  We observe that the usual static wormhole geometry is obtained as a special case of the above general solution for $\beta=0$ and $a(t)=constant.$ Thus we see that the above general metric solution can describe a variety of spatial geometry depending upon the choice of parameters.

To provide some viable dynamic wormhole solutions we give the following examples for $g(r).$

\begin{description}
\item[$\blacksquare$] Here we consider $\alpha=constant.$ 
This gives $g(r)=\left(\frac{r}{r_{0}}\right)^{\frac{1}{\alpha}}$ with $r_{0}$ being the constant of integration. Hence we obtain the modified throat function as:
\begin{equation}
\frac{b_{n}(r)}{r}=\left(\frac{r}{r_{0}}\right)^{\frac{2}{\alpha}}\label{b1}
\end{equation}
which will define a viable wormhole throat for $\alpha<0$ \cite{sb1},   \cite{sb2}: 

\item[$\blacksquare$] Here we consider $g(r)=\left(\frac{r}{r_{0}}\right)^{n/2},~n<0,$ with $r_{0}$ being a constant. This gives the throat function as \cite{sb2}, \cite{tb}:
\begin{equation}
\frac{b_{n}(r)}{r}=\left(\frac{r}{r_{0}}\right)^{n}\label{b2}.
\end{equation}

\item[$\blacksquare$] Next we consider $g(r)=\sqrt{\frac{1}{1+r-r_{0}}}.$ This defines the wormhole throat as \cite{sb2}:
\begin{equation}
\frac{b_{n}(r)}{r}=\frac{1}{1+r-r_{0}}\label{b3}
\end{equation}

\item[$\blacksquare$] Another viable throat can be: $g(r)=\sqrt{C\left(1-\frac{r_{0}}{r}\right)+\frac{r_{0}}{r}}$ for $C\leq 1.$ This will describe the corresponding shape function as \cite{tb}:
\begin{equation}
\frac{b_{n}(r)}{r}=C\left(1-\frac{r_{0}}{r}\right)+\frac{r_{0}}{r}\label{b4}.
\end{equation}
\end{description}

One can easily verify that all of the above examples defines viable wormhole throat, and that $b_{n}(r_{0})=r_{0}$ where $r_{0}$ is the wormhole throat. 

Thus starting from the most general wormhole metric we have successfully described several possible evolving wormhole solutions. Next we will find the nature of the matter stress energy tensor that can support the above systems. We will also provide energy conditions satisfied by such matter stress energy tensors. In the following section we analyse the energy conditions in some details.

\section{energy Conditions}

In order to find the nature of the matter stress energy supporting the wormhole, we consider the equation (\ref{consv2}) coupled with the relation $p_{r}(r,t)=\alpha p_{t}(r,t).$ This gives: 
\begin{align}
p_{r}(r,t)&=p_{0}(t)\left(\frac{g(r)}{r}\right)^{2}\label{pr}\\
p_{t}(r,t)&=p_{0}(t)\left(\frac{g(r)}{r}\right)g'(r)\label{pt}
\end{align}
where $p_{0}(t)$ is an arbitrary function of $t$ and is obtained as $p_{0}(t)=-\frac{1}{a^{2}(t)}$ from the equations (\ref{epr}) and (\ref{ept}) after substituting the values of $b(r)$ and $a(t).$
The value of the energy component of matter is evaluated as:
\begin{equation}
\rho(r,t)=3H^{2}e^{-2\phi(t)}-\frac{3\beta}{a^{2}}+\frac{2g(r)g'(r)}{a^{2}r}+\frac{1}{a^{2}}\left(\frac{g(r)}{r}\right)^{2}\label{rho}
\end{equation}
Accordingly we obtain $\rho+p_{r}=3H^{2}e^{-2\phi(t)}-\frac{3\beta}{a^{2}}+\frac{2g(r)g'(r)}{a^{2}r}$ and $\rho+p_{t}=3H^{2}e^{-2\phi(t)}-\frac{3\beta}{a^{2}}+\frac{g(r)g'(r)}{a^{2}r}+\frac{1}{a^{2}}\left(\frac{g(r)}{r}\right)^{2}.$ This shows that the NEC depends on the parameters $\beta$ and $\alpha.$

\subsection{Energy conditions at the throat for trivial $\phi(t)$}

Corresponding to the choice of $\phi(t)$ given by $a(t)=e^{\phi(t)}$ the energy conditions for $\beta=\pm 1$ are
\begin{align}
\rho+p_{r}=2\frac{g(r)g'(r)}{ra^{2}}\\
\rho+p_{t}=\frac{g(r)g'(r)}{ra^{2}}+\frac{g^{2}(r)}{(ar)^{2}}.
\end{align}

Hence for trivial $\phi$ the energy conditions effectively reduce to $\frac{gg'}{a^{2}}>0$ for both $\beta=\pm 1,$ which shows that at the throat, the energy conditions are actually dependent on the properties of the shape function and the parameter $\alpha(r).$\\
For the choice of $\beta=0$ we can find that the energy conditions are given by:
\begin{align}
\rho+p_{r}=\frac{1}{a^{2}}\left(\frac{12}{t^{2}}+2\frac{g(r)g'(r)}{r}\right)\\
\rho+p_{t}=\frac{1}{a^{2}}\left(\frac{12}{t^{2}}+\frac{g(r)g'(r)}{r}+\frac{g^{2}(r)}{r^{2}}\right)
\end{align}
In this case the energy conditions will depend upon the location of the wormhole throat at any given finite slice of time $t.$ Table 1, gives a tabular description of the null energy conditions corresponding to the several viable choices of the shape function cited above.

From the table one can see that for conditions of $\beta\neq 0$ the wormhole would essentially require NEC violation for any arbitrary time slice $t.$ For $\beta\geq 0$ as $t\rightarrow\infty,~\rho+p_{r},~\rho+p_{t}$ both can marginally satisfy the NEC. This is because at large values of $t$ the matter-energy tensor becomes negligible in such wormholes. For values of $\beta=0,$ at any arbitrary time slice $t=t_{0}$ one can get wormholes with energy conditions satisfied provided the throat satisfies certain parameter restrictions as listed in Table 1. For such wormholes NEC will be satisfied for all time slices only if those parameters vary with time subject to the conditions stated.

\subsection{Energy conditions at the throat for any general $\phi(t)$} 

For the case $\beta=0$ we found the solution as $a(t)=a_{0}(\int e^{\phi(t)}dt)^{\frac{2}{3}},$ corresponding to any $\phi(t).$ As a specific example we consider the case where $e^{\phi(t)}=\hat{\phi}+\left(\frac{t}{t_{0}}\right)^{m},~m<-1$ and $\hat{\phi}$ a finite constant quantity such that $\hat{\phi}\neq -\frac{1}{1+m}$.  Accordingly the solution for $a(t)$ is obtained as: $a(t)=a_{0}\left[\hat{\phi}t+\frac{t^{m+1}}{(m+1)t_{0}^{m}}\right]^{2/3},$ with $a_{0}$ being some arbitrary constant. With this $a(t)$ and $\phi(t),$ the energy conditions are:
\begin{align}
\rho+p_{r}=\frac{1}{a^{2}}\left[\frac{4a_{0}^{5}}{3a}+\frac{2g(r)g'(r)}{r}\right]\\
\rho+p_{t}=\frac{1}{a^{2}}\left[\frac{4a_{0}^{5}}{3a}+\frac{g(r)g'(r)}{r}+\frac{g^{2}(r)}{r^{2}}\right]
\end{align}
As $t\rightarrow\pm\infty,~\frac{1}{a}\rightarrow 0$ it is observed that both $\rho+p_{r}$ and $\rho+p_{t}$ approach zero. Further for any finite time slice $t$ the NEC depends on the location of the wormhole throat at that instant of time, which again is restricted by its corresponding parameters. Table 2 gives the tabular description of the various shape functions and the corresponding parameter restrictions for validity of NEC.

Finally we consider the energy conditions at the wormhole throat corresponding to the solution $a(t)=(c_{1}t+c_{2})^{2/3}$ and $e^{2\phi(t)}=\frac{4c_{1}^{2}}{9\beta}(c_{1}t+c_{2})^{-1/3}.$ The energy conditions are obtained as:
\begin{align}
\rho+p_{r}=\frac{3\beta}{a^{2}}\left[(c_{1}t+c_{2})^{-1/3}-1\right]+\frac{2g(r)g'(r)}{ra^{2}}\\
\rho+p_{t}=\frac{3\beta}{a^{2}}\left[(c_{1}t+c_{2})^{-1/3}-1\right]+\frac{g(r)g'(r)}{a^{2}r}+\frac{1}{a^{2}}\left(\frac{g(r)}{r}\right)^{2}.
\end{align}
It is observed that for $\beta\leq 0$ the energy conditions at the throat are always violated for any finite time slice $t.$ While for $\beta>0$ we can possibly obtain wormholes with NEC subject to certain parameter restrictions as shown in Table 3. The criterion mentioned in Table 3 are all valid only for choice of $c_{1}$ and $c_{2}$ such that $(c_{1}t+c_{2})^{-1/3}>1.$

\subsection{Energy Conditions Corresponding to the Static Case}

In the static case, we show that the NEC is violated for the above suggested wormholes. In this case, the expressions of $\rho+p_r$ and $\rho+p_t$ change to $\rho+p_r=\frac{2g(r)g'(r)}{r}$ and
$\rho+p_t=\frac{g(r)g'(r)}{r}+\frac{g^2(r)}{r^2}$. So 

a. Let $g^2(r)=\left(\frac{r}{r_0}\right)^{2/\alpha}$: In this
case $\rho+p_r=\frac{2}{\alpha {r_0}^2}$ with $\alpha<0$.\\

b. Let $g^2(r)=\left(\frac{r}{r_0}\right)^{n}$: In this case
we get $\rho+p_r=\frac{n}{{r_0}^2}$ with $n<0$.\\

c. Let $g^2(r)=\frac{1}{1+r-r_0}$: For this example we get
$\rho+p_r=-\frac{1}{r_0}$.\\

d. Let $g^2(r)=c\left(1-\frac{r_0}{r}\right)+\frac{r_0}{r}$. In
this case $\rho+p_r=-\frac{1-c}{{r_0}^2}$ with $c<1$.\\

Which shows that NEC is violated in all the above examples as in the case of static wormholes. This clearly establishes that in the case of evolving wormholes, the time dependence helps in temporary suspension of NEC. 

\begin{center}
\begin{table}
\begin{tabular}{|l|l p{6cm}|l|}
\hline
$g^{2}(r)$&\multicolumn{2}{|c|}{Energy Conditions at $r=r_{0}$}&NEC\\
\hline
\multirow{2}{*}{$\left(\frac{r}{r_{0}}\right)^{\frac{2}{\alpha}},~\alpha<0$}&$\beta\neq 0:$ &$
\rho+p_{r}=\frac{2}{\alpha(r_{0}a)^{2}}$

$\rho+p_{t}=\frac{1}{(r_{0}a)^{2}}\left(1+\frac{1}{\alpha}\right).$&Violated \\
\cline{2-4}
&$\beta=0:$ &$
\rho+p_{r}=\frac{12}{(at)^{2}}+\frac{2}{\alpha(r_{0}a)^{2}}$

$\rho+p_{t}=\frac{12}{(at)^{2}}+\frac{1}{(r_{0}a)^{2}}\left(1+\frac{1}{\alpha}\right).$& Satisfied for $\alpha<-\frac{t_{0}^{2}}{6r_{0}^{2}}$ \\
\hline 
\multirow{2}{*}{$\left(\frac{r}{r_{0}}\right)^{n},~n<0$}&$\beta\neq 0:$ &$
\rho+p_{r}=\frac{n}{\alpha(r_{0}a)^{2}}$

$\rho+p_{t}=\frac{1}{(r_{0}a)^{2}}\left(1+\frac{n}{2}\right).$&Violated \\
\cline{2-4}
&$\beta=0:$ &$
\rho+p_{r}=\frac{12}{(at)^{2}}+\frac{n}{(r_{0}a)^{2}}$

$\rho+p_{t}=\frac{12}{(at)^{2}}+\frac{1}{(r_{0}a)^{2}}\left(1+\frac{n}{2}\right).$&Satisfied for $n>-\frac{12r_{0}^{2}}{t_{0}^{2}}$ \\
\hline
\multirow{2}{*}{$\left(\frac{1}{1+r-r_{0}}\right)$}&$\beta\neq 0:$ &$
\rho+p_{r}=-\frac{1}{r_{0}a^{2}}$

$\rho+p_{t}=\frac{1}{r_{0}a^{2}}\left(-\frac{1}{2}+\frac{1}{r_{0}}\right).$&Violated \\
\cline{2-4}
&$\beta=0:$ &$
\rho+p_{r}=\frac{12}{(at)^{2}}-\frac{1}{r_{0}a^{2}}$

$\rho+p_{t}=\frac{12}{(at)^{2}}+\frac{1}{r_{0}a^{2}}\left(-\frac{1}{2}+\frac{1}{r_{0}}\right).$&Satisfied for $\frac{12r_{0}}{t_{0}^{2}}>1$ \\
\hline
\multirow{2}{*}{$c\left(1-\frac{r_{0}}{r}\right)+\frac{r_{0}}{r},~c<1$}&$\beta\neq 0:$ &$
\rho+p_{r}=-\frac{1-c}{(r_{0}a)^{2}}$

$\rho+p_{t}=\frac{1}{(r_{0}a)^{2}}\left(1+\frac{c-1}{2}\right).$&Violated \\
\cline{2-4}
&$\beta=0:$ &$
\rho+p_{r}=\frac{12}{(at)^{2}}-\frac{1-c}{(r_{0}a)^{2}}$

$\rho+p_{t}=\frac{12}{(at)^{2}}+\frac{1}{(r_{0}a)^{2}}\left(1+\frac{c-1}{2}\right).$&Satisfied for $c>1-\frac{12r_{0}^{2}}{t_{0}^{2}}$\\
\hline
\end{tabular}
\caption{The Energy Conditions for $a(t)=e^{\phi(t)}$ with different shape functions at time slice $t=t_{0}.$ }
\end{table}
\end{center}

\begin{center}
\begin{table}
\begin{tabular}{|l|p{6cm}|l|}
\hline
$g^{2}(r)$&Energy Conditions at $r=r_{0}$&NEC\\
\hline
$\left(\frac{r}{r_{0}}\right)^{\frac{2}{\alpha}},~\alpha<0$&$
\rho+p_{r}=\frac{1}{a^{2}}\left[\frac{4a_{0}^{5}}{3a}+\frac{2}{\alpha r_{0}^{2}}\right]$

$\rho+p_{t}=\frac{1}{a^{2}}\left[\frac{4a_{0}^{5}}{3a}+\frac{1}{r_{0}^{2}}\left(1+\frac{1}{\alpha}\right)\right].$& Satisfied for $\alpha<-\frac{3t_{0}^{2/3}}{2r_{0}^{2}a_{0}^{4}}\left(\hat{\phi}+\frac{1}{m+1}\right)^{2/3}$ \\
\hline 
$\left(\frac{r}{r_{0}}\right)^{n},~n<0$&$
\rho+p_{r}=\frac{1}{a^{2}}\left[\frac{4a_{0}^{5}}{3a}+\frac{n}{r_{0}^{2}}\right]$

$\rho+p_{t}=\frac{1}{a^{2}}\left[\frac{4a_{0}^{5}}{3a}+\frac{1}{r_{0}^{2}}\left(1+\frac{n}{2}\right)\right].$&Satisfied for $n>-\frac{4r_{0}^{2}a_{0}^{4}t_{0}^{-2/3}}{3\left(\hat{\phi}+\frac{1}{m+1}\right)^{2/3}}$ \\
\hline
$\left(\frac{1}{1+r-r_{0}}\right)$&$
\rho+p_{r}=\frac{1}{a^{2}}\left[\frac{4a_{0}^{5}}{3a}-\frac{1}{r_{0}}\right]$

$\rho+p_{t}=\frac{1}{a^{2}}\left[\frac{4a_{0}^{5}}{3a}+\frac{1}{r_{0}}\left(-\frac{1}{2}+\frac{1}{r_{0}}\right)\right].$&Satisfied for $\frac{4r_{0}a_{0}^{4}}{3t_{0}^{2/3}\left(\hat{\phi}+\frac{1}{m+1}\right)^{2/3}}>1$ \\
\hline
$c\left(1-\frac{r_{0}}{r}\right)+\frac{r_{0}}{r},~c<1$&$
\rho+p_{r}=\frac{1}{a^{2}}\left[\frac{4a_{0}^{5}}{3a}-\frac{1-c}{r_{0}^{2}}\right]$

$\rho+p_{t}=\frac{1}{a^{2}}\left[\frac{4a_{0}^{5}}{3a}+\frac{1}{r_{0}^{2}}\left(1-\frac{1-c}{2}\right)\right].$&Satisfied for $c>1-\frac{4r_{0}^{2}a_{0}^{4}}{3t_{0}^{2/3}\left(\hat{\phi}+\frac{1}{m+1}\right)^{2/3}}$\\
\hline
\end{tabular}
\caption{The Energy Conditions for $a(t)=a_{0}\left[\hat{\phi}t+\frac{t^{m+1}}{(m+1)t_{0}^{m}}\right]^{2/3}$ with different shape functions at time slice $t=t_{0}.$}
\end{table}
\end{center}

\section{Graphical Representation}

With the help of numerical evaluation we graphically represent the energy conditions corresponding to the parameter restrictions as mentioned in tables 1, 2 and 3. For every figure we have considered the variation of a time slice $t_{0}$ and a corresponding variation of the location of the throat. We have also shown how the energy conditions behave corresponding to each of the shape functions as given by (\ref{b1}), (\ref{b2}), (\ref{b3}) and (\ref{b4}). All the figures are plotted subject to parameter restrictions in regions of validity of the NEC. Figure 1 correspond to scale factor $a(t)=e^{\phi(t)}$ and for (\ref{b1}), (\ref{b2}) and (\ref{b4}). Figure 2 show the values of $\rho+p_{r}$ and $\rho+p_{t}$ for $a(t)= a_{0}\left[\hat{\phi}t+\frac{t^{m+1}}{(m+1)t_{0}^{m}}\right]^{2/3}.$ The numerical evaluation of the energy conditions and correspondingly the figures were generated for arbitrary values of parameters $a_{0},~\hat{\phi}$ and $m.$ Similar representative figures could be obtained for any other values of the parameter, subject to the constraint. Figure 3 is for $a(t)=(c_{1}t+c_{2})^{2/3}$ corresponding to $\beta=1.$ Here again the figures were generated for arbitrary values of the parameters $c_{1}$ and $c_{2}$ subject to the tabulated constraints and an additional requirement of $(c_{1}t+c_{2})^{-1/3}>1$. Figure 4 is for shape function given by (\ref{b3}) for the three different scale factors discussed above. Evidently evolving wormhole geometries discussed here can exist for some arbitrary finite amount of time without violating the NEC. However such wormhole throat will have continuous temporal dependence subject to parameter constraints, hence NEC will only hold at some arbitrary finite time instants where the constraints are met.

\begin{center}
\begin{table}
\begin{tabular}{|l|p{6cm}|l|}
\hline
$g^{2}(r)$&Energy Conditions at $r=r_{0}$&NEC\\
\hline
$\left(\frac{r}{r_{0}}\right)^{\frac{2}{\alpha}},~\alpha<0$&$
\rho+p_{r}=\frac{3}{a^{2}}\left[(c_{1}t+c_{2})^{-1/3}-1\right]+\frac{2}{\alpha r_{0}^{2}a^{2}}$

$\rho+p_{t}=\frac{3}{a^{2}}\left[(c_{1}t+c_{2})^{-1/3}-1\right]+\frac{1}{r_{0}^{2}a^{2}}\left(1+\frac{1}{\alpha}\right).$& Satisfied for $\alpha<-\frac{2}{3r_{0}^{2}\left[(c_{1}t_{0}+c_{2})^{-1/3}-1\right]}$ \\
\hline 
$\left(\frac{r}{r_{0}}\right)^{n},~n<0$&$
\rho+p_{r}=\frac{3}{a^{2}}\left[(c_{1}t+c_{2})^{-1/3}-1\right]+\frac{n}{r_{0}^{2}a^{2}}$

$\rho+p_{t}=\frac{3}{a^{2}}\left[(c_{1}t+c_{2})^{-1/3}-1\right]+\frac{1}{r_{0}^{2}a^{2}}\left(1+\frac{n}{2}\right).$&Satisfied for $n>-3r_{0}^{2}\left[(c_{1}t_{0}+c_{2})^{-1/3}-1\right]$ \\
\hline
$\left(\frac{1}{1+r-r_{0}}\right)$&$
\rho+p_{r}=\frac{3}{a^{2}}\left[(c_{1}t+c_{2})^{-1/3}-1\right]-\frac{1}{r_{0}a^{2}}$

$\rho+p_{t}=\frac{3}{a^{2}}\left[(c_{1}t+c_{2})^{-1/3}-1\right]+\frac{1}{r_{0}a^{2}}\left(-\frac{1}{2}+\frac{1}{r_{0}}\right).$&Satisfied for $3r_{0}\left[(c_{1}t_{0}+c_{2})^{-1/3}-1\right]>1$ \\
\hline
$c\left(1-\frac{r_{0}}{r}\right)+\frac{r_{0}}{r},~c<1$&$
\rho+p_{r}=\frac{3}{a^{2}}\left[(c_{1}t+c_{2})^{-1/3}-1\right]-\frac{1-c}{r_{0}^{2}a^{2}}$

$\rho+p_{t}=\frac{3}{a^{2}}\left[(c_{1}t+c_{2})^{-1/3}-1\right]+\frac{1}{r_{0}^{2}a^{2}}\left(1-\frac{1-c}{2}\right).$&Satisfied for $c>1-3r_{0}^{2}\left[(c_{1}t_{0}+c_{2})^{-1/3}-1\right]$\\
\hline
\end{tabular}
\caption{The Energy Conditions for $a(t)=(c_{1}t+c_{2})^{2/3}$ for $\beta=1$ with different shape functions at time slice $t=t_{0}.$}
\end{table}
\end{center}

\begin{figure}[htp]
\centering
\includegraphics[width=.25\textwidth]{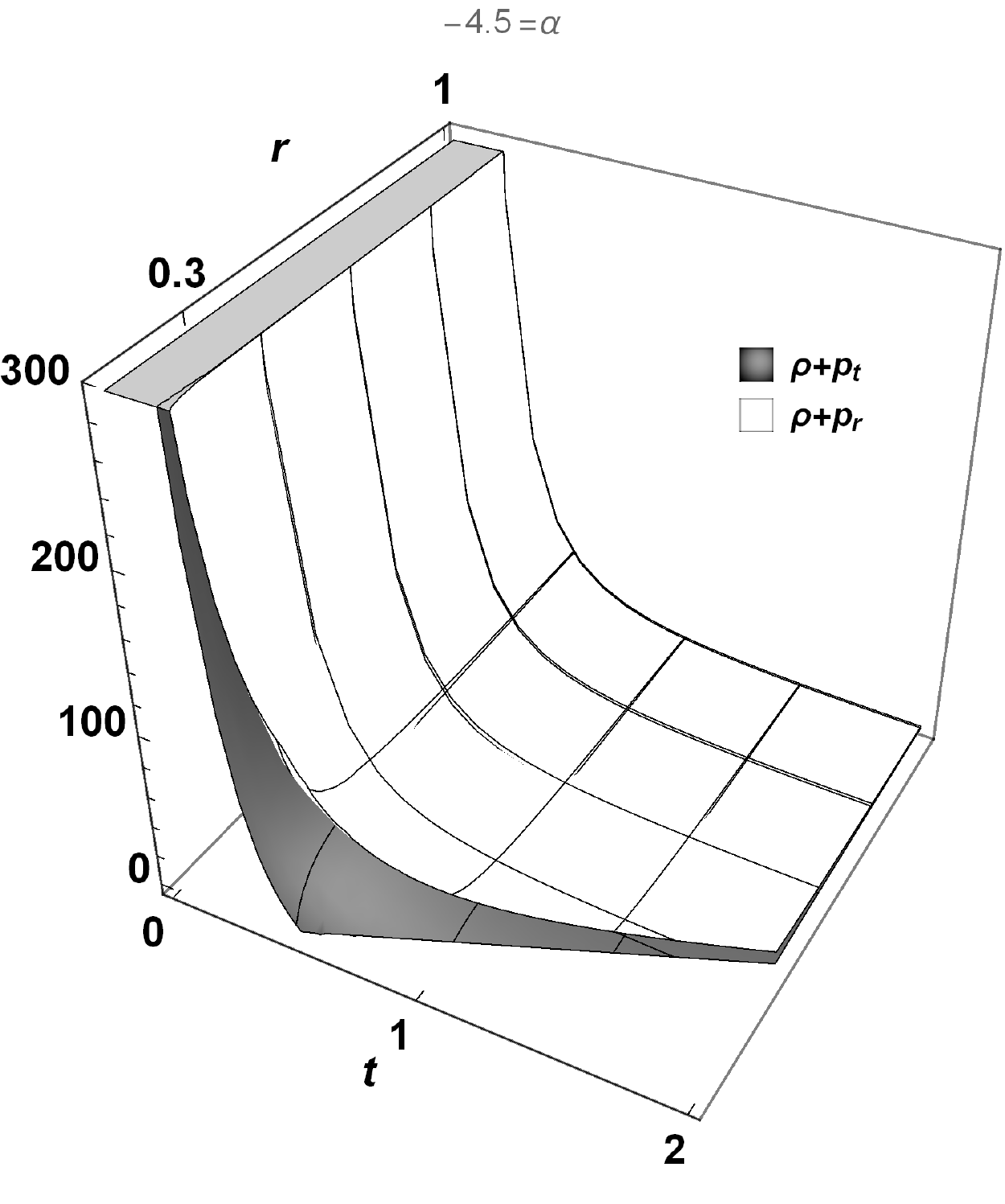}\quad
\includegraphics[width=.25\textwidth]{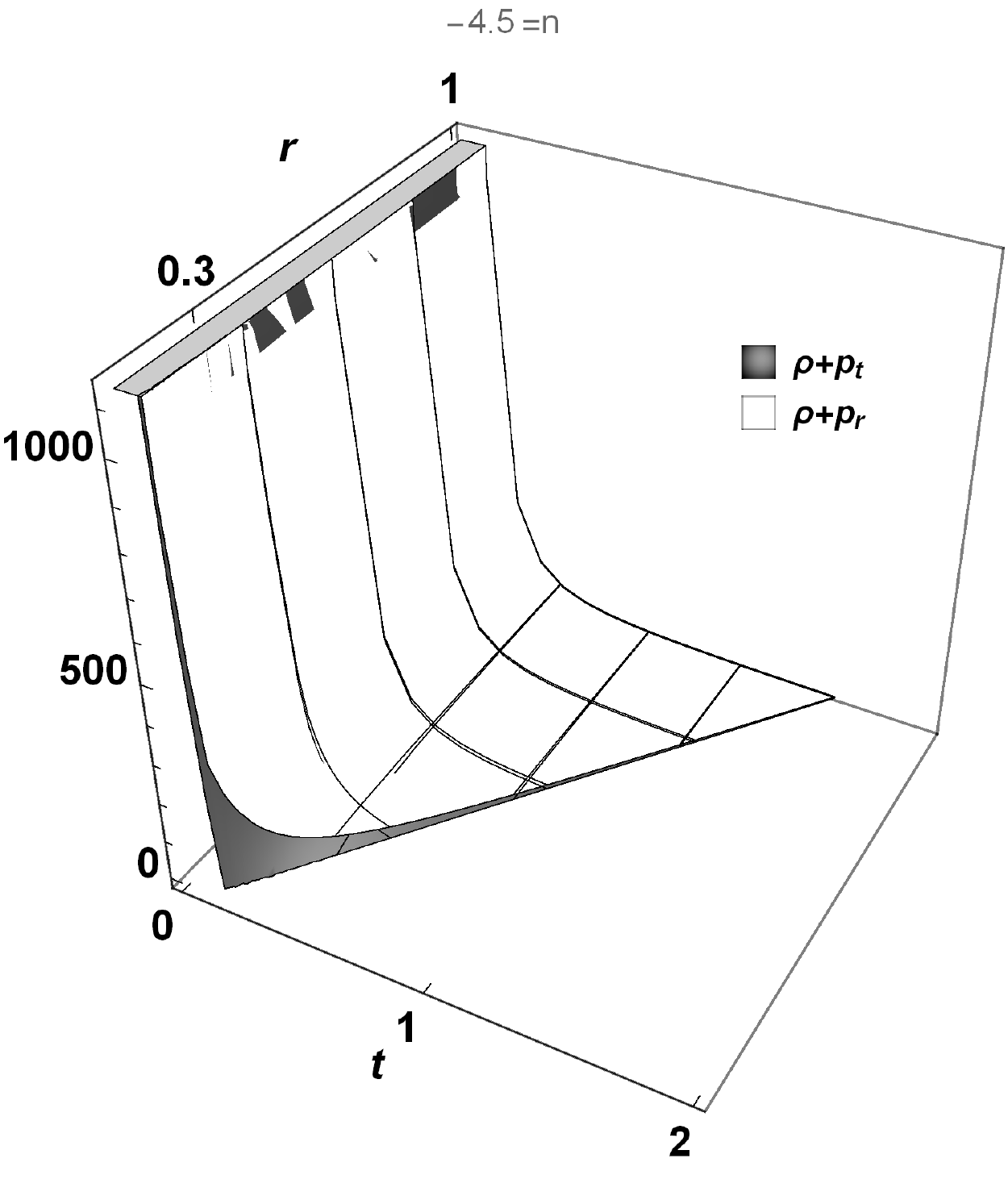}\quad
\includegraphics[width=.25\textwidth]{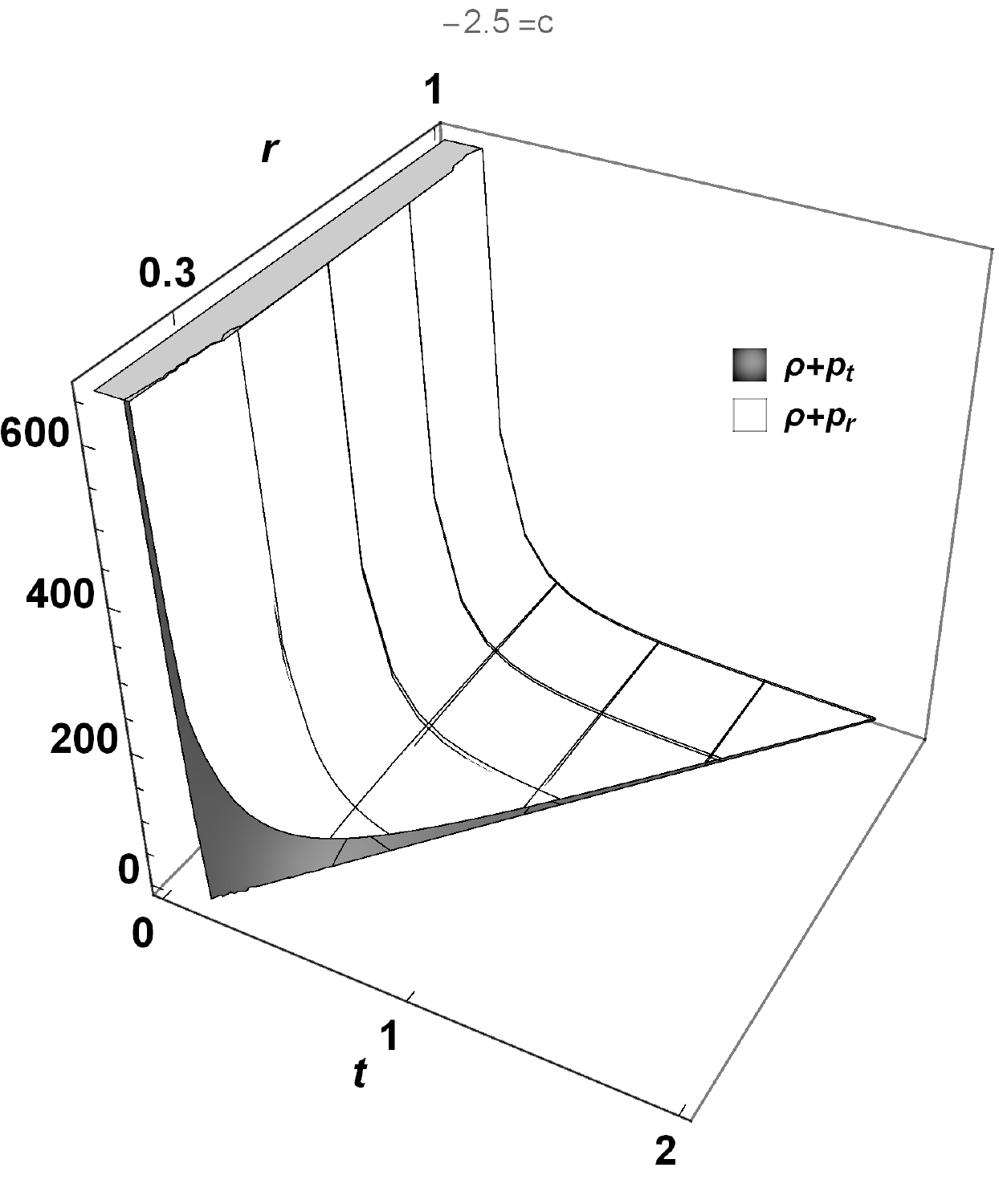}

\caption{The above panel (from left to right in order) show the variations in $\rho+ p_{r}$ and $\rho+ p_{t}$ for a) $g^{2}(r)=\left(\frac{r}{r_{0}}\right)^{\frac{2}{\alpha}},$ subject to the condition that $\alpha<-\frac{t_{0}^{2}}{6r_{0}^{2}}$ b) $g^{2}(r)=\left(\frac{r}{r_{0}}\right)^{n},$ subject to the condition that $n>-\frac{12r_{0}^{2}}{t_{0}^{2}}$ c)$g^{2}(r)=c\left(1-\frac{r_{0}}{r}\right)+\frac{r_{0}}{r},$ subject to the condition that $c>1-\frac{12r_{0}^{2}}{t_{0}^{2}}$ corresponding to $a(t)=e^{\phi(t)}.$}
\end{figure}

\begin{figure}[htp]
\centering
\includegraphics[width=.25\textwidth]{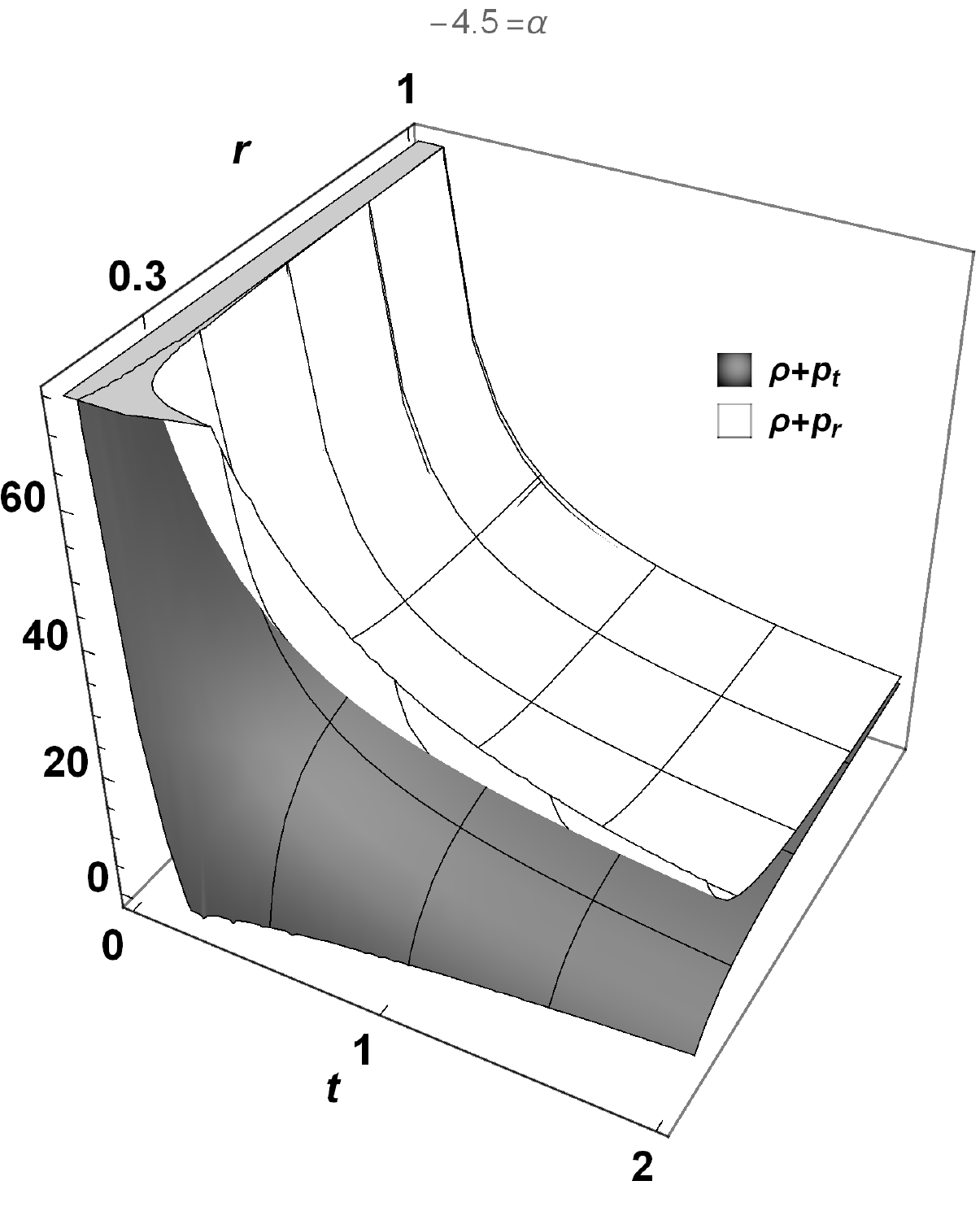}\quad
\includegraphics[width=.25\textwidth]{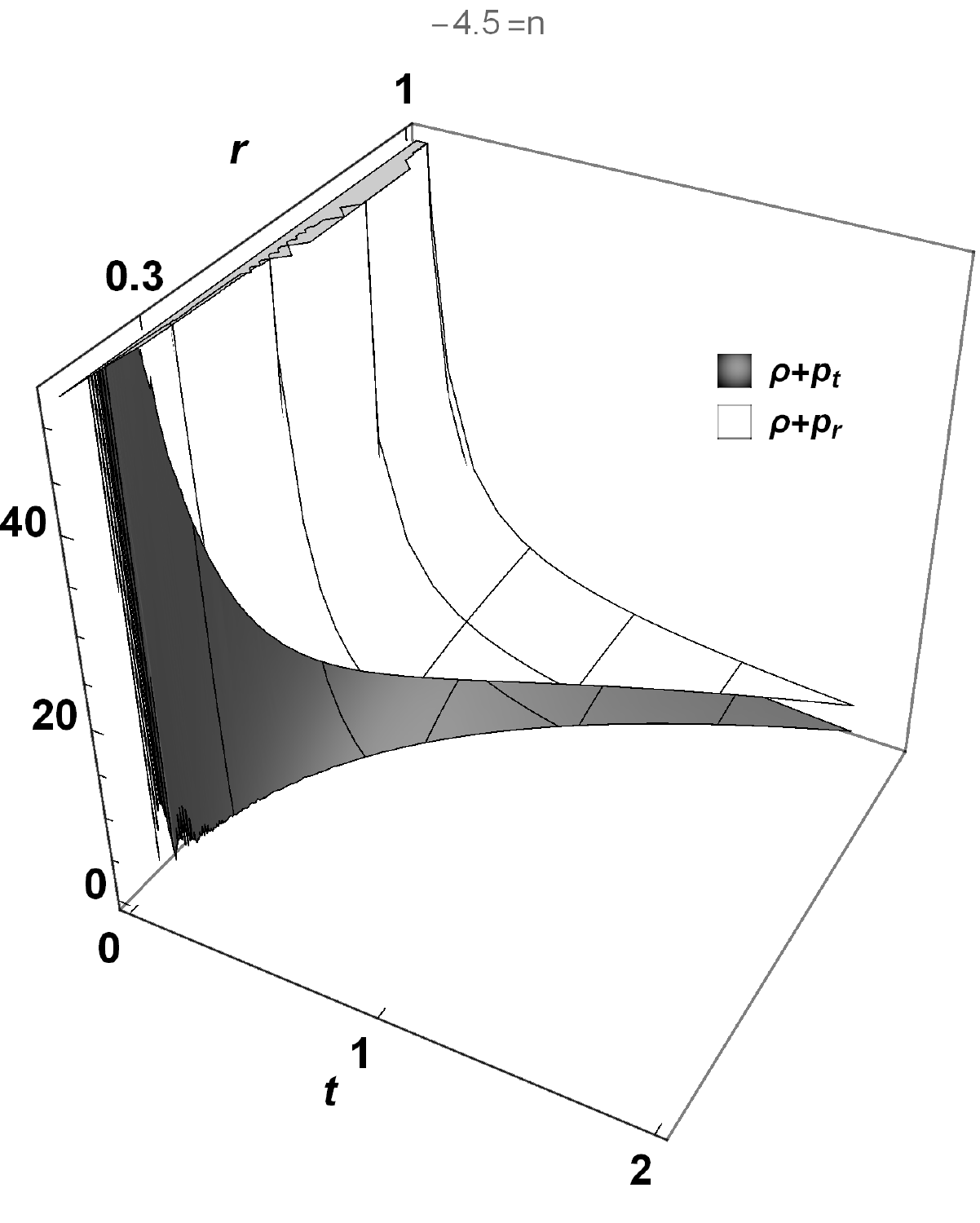}\quad
\includegraphics[width=.25\textwidth]{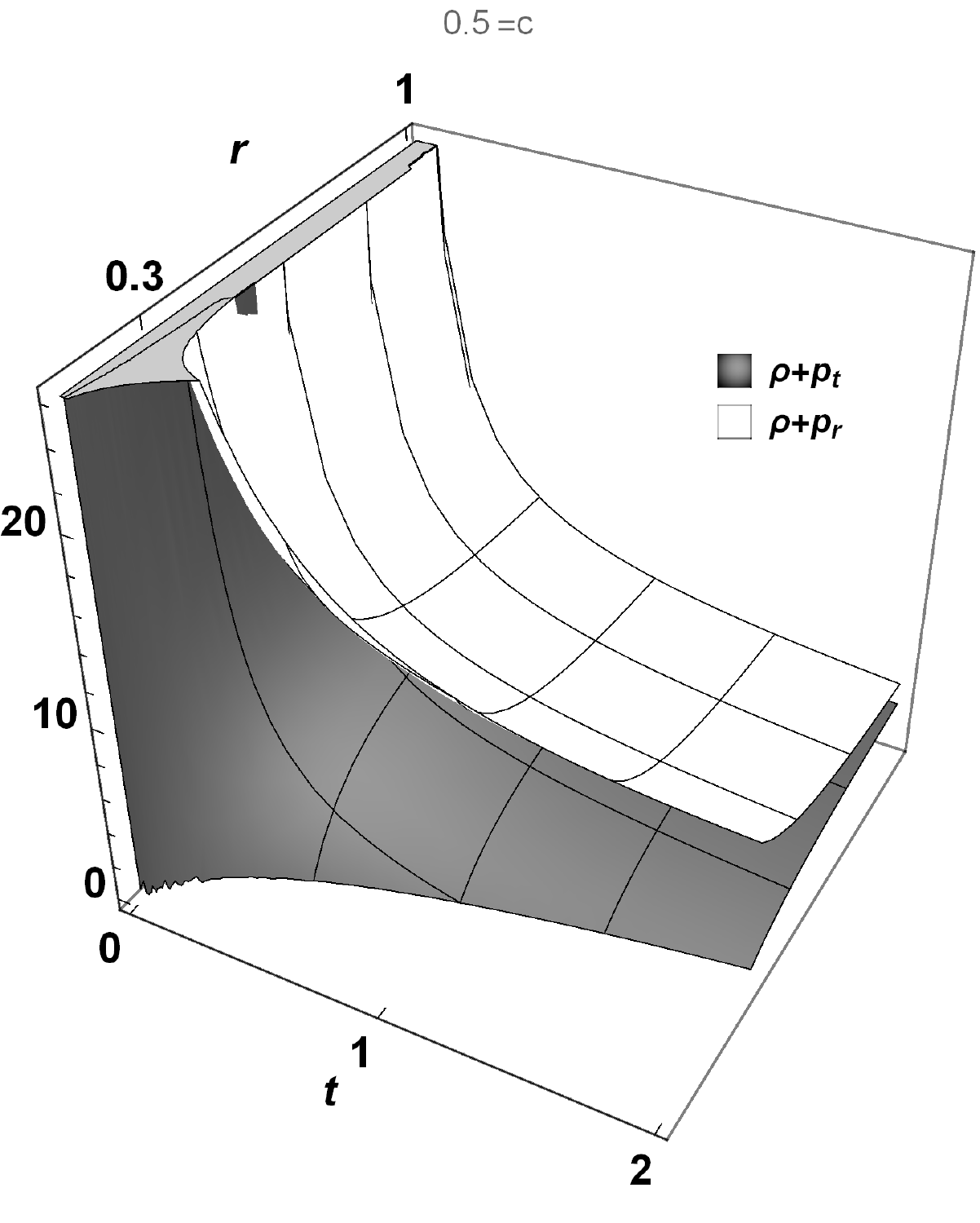}

\caption{The above panel (from left to right in order) show the variations in $\rho+ p_{r}$ and $\rho+ p_{t}$ for a) $g^{2}(r)=\left(\frac{r}{r_{0}}\right)^{\frac{2}{\alpha}},$ subject to the condition that $\alpha<-\frac{3t_{0}^{2/3}}{2r_{0}^{2}a_{0}^{4}}\left(\hat{\phi}+\frac{1}{m+1}\right)^{2/3}.$ The parameters taken are $a_{0} = 1.8;~\hat{\phi} = 1.5;~m = -2.4.$ b) $g^{2}(r)=\left(\frac{r}{r_{0}}\right)^{n},$ subject to the condition that $n>-\frac{4r_{0}^{2}a_{0}^{4}t_{0}^{-2/3}}{3\left(\hat{\phi}+\frac{1}{m+1}\right)^{2/3}}.$ The parameters taken are: $a_{0} = 1.5;~\hat{\phi} = 1.5;~m = -3.$ c) $g^{2}(r)=c\left(1-\frac{r_{0}}{r}\right)+\frac{r_{0}}{r},$ subject to the condition that $c>1-\frac{4r_{0}^{2}a_{0}^{4}}{3t_{0}^{2/3}\left(\hat{\phi}+\frac{1}{m+1}\right)^{2/3}}.$ The parameters used are $a_{0} = 1.5;~\hat{\phi} = 2.2;~m = -3.$ Here $a(t)=a_{0}\left[\hat{\phi}t+\frac{t^{m+1}}{(m+1)t_{0}^{m}}\right]^{2/3}.$}
\end{figure}

\begin{figure}[htp]
\centering
\includegraphics[width=.25\textwidth]{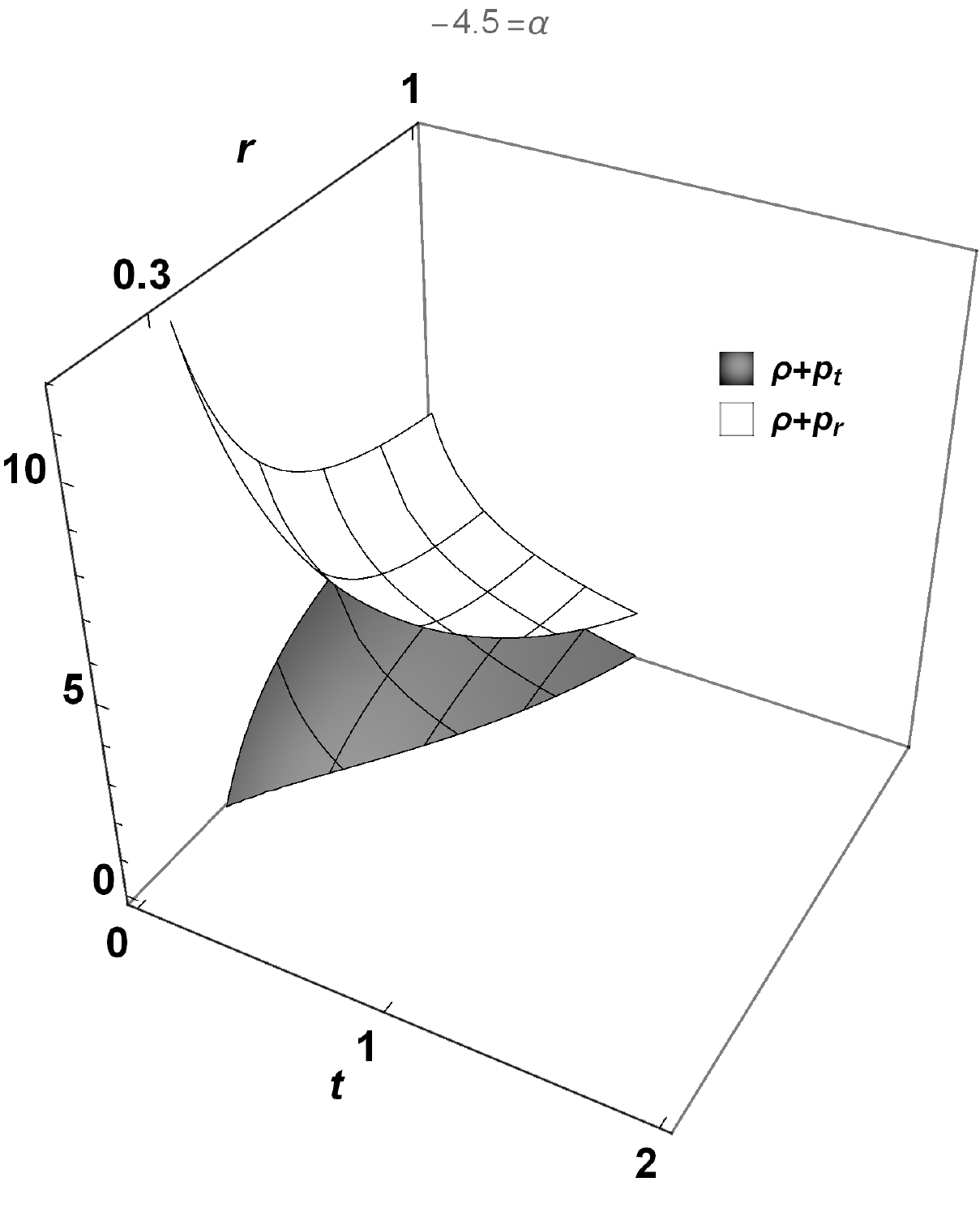}\quad
\includegraphics[width=.25\textwidth]{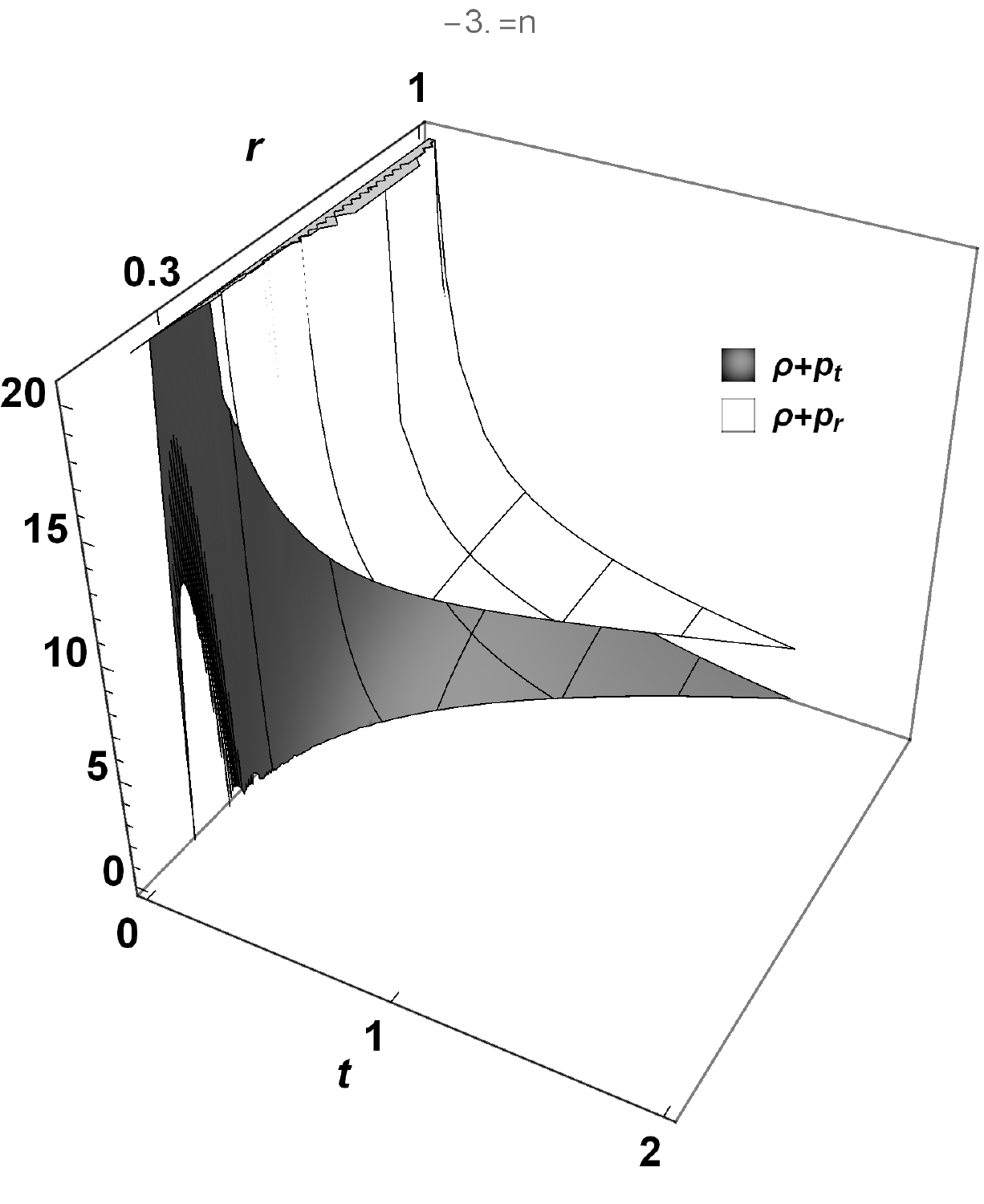}\quad
\includegraphics[width=.25\textwidth]{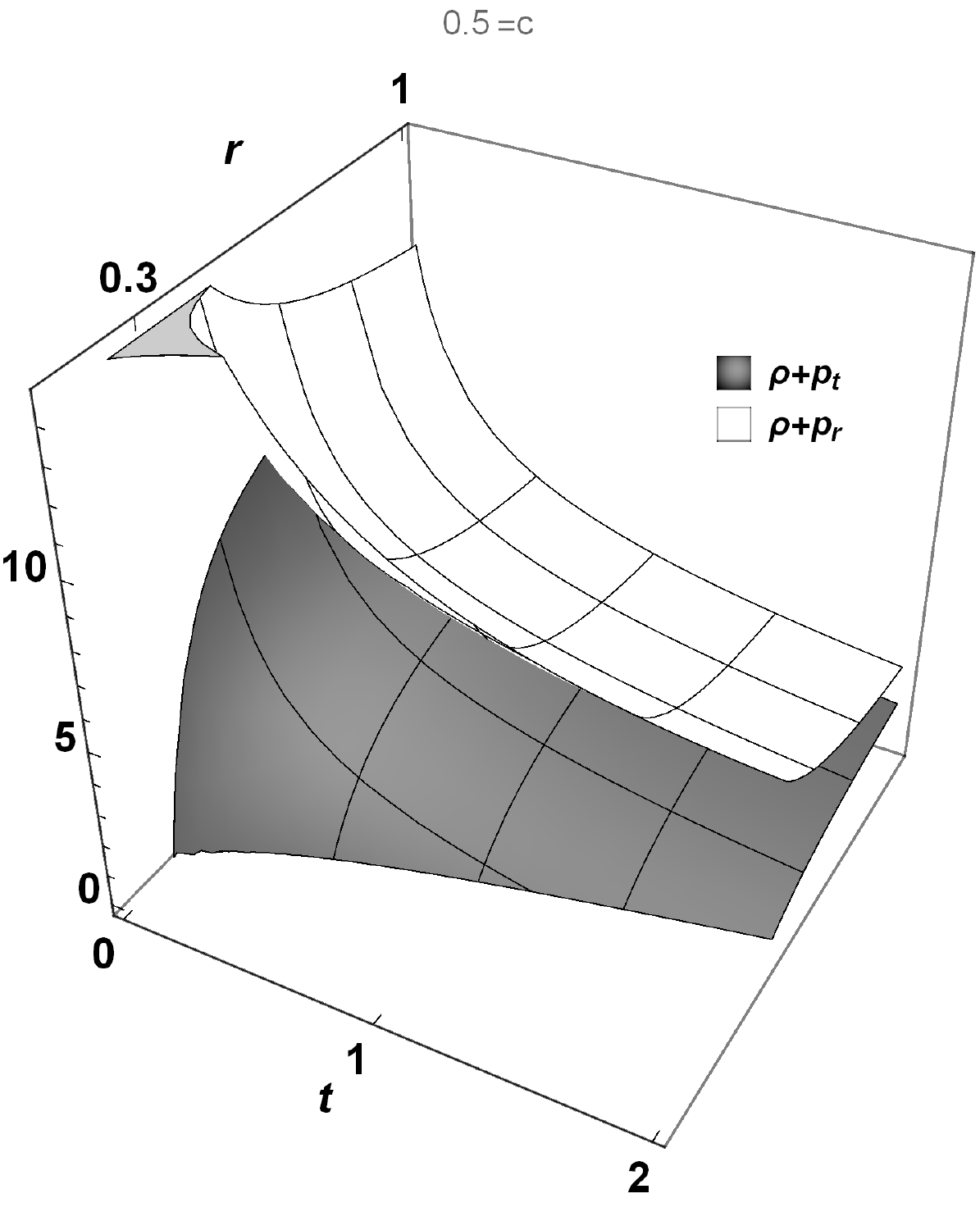}

\caption{The above panels (from left to right in order) show the variation in $\rho+ p_{r}$ and $\rho+ p_{t}$ corresponding to a) $g^{2}(r)=\left(\frac{r}{r_{0}}\right)^{\frac{2}{\alpha}},$ subject to the condition that $\alpha<-\frac{2}{3r_{0}^{2}\left[(c_{1}t_{0}+c_{2})^{-1/3}-1\right]}.$ The parameters taken are $c_{1} = 0.65;~c_{2} = 0.07.$, b) $g^{2}(r)=\left(\frac{r}{r_{0}}\right)^{n},$ subject to the condition that $n>-\frac{4r_{0}^{2}a_{0}^{4}t_{0}^{-2/3}}{3\left(\hat{\phi}+\frac{1}{m+1}\right)^{2/3}}.$ The parameters taken are: $c_{1} = .08;~c_{2} = 0.0$, c) $g^{2}(r)=c\left(1-\frac{r_{0}}{r}\right)+\frac{r_{0}}{r},$ subject to the codition that $c>1-\frac{4r_{0}^{2}a_{0}^{4}}{3t_{0}^{2/3}\left(\hat{\phi}+\frac{1}{m+1}\right)^{2/3}}.$ The parameters used are $c_{1} = 0.1;~c_{2} = 0.01.$ Here $a(t)=(c_{1}t+c_{2})^{2/3}.$}
\end{figure}

\begin{figure}[htp]
\centering
\includegraphics[width=.25\textwidth]{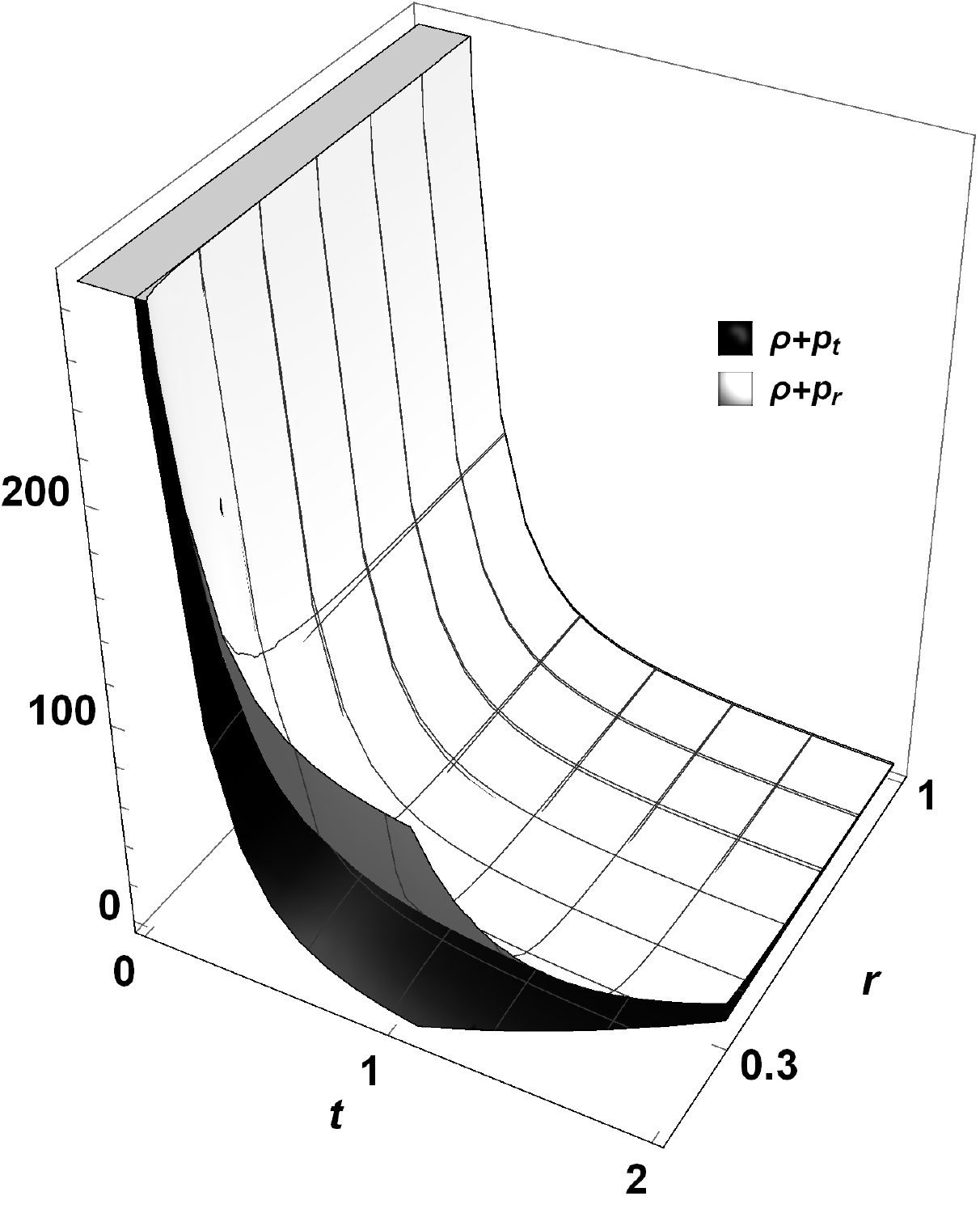}\quad
\includegraphics[width=.25\textwidth]{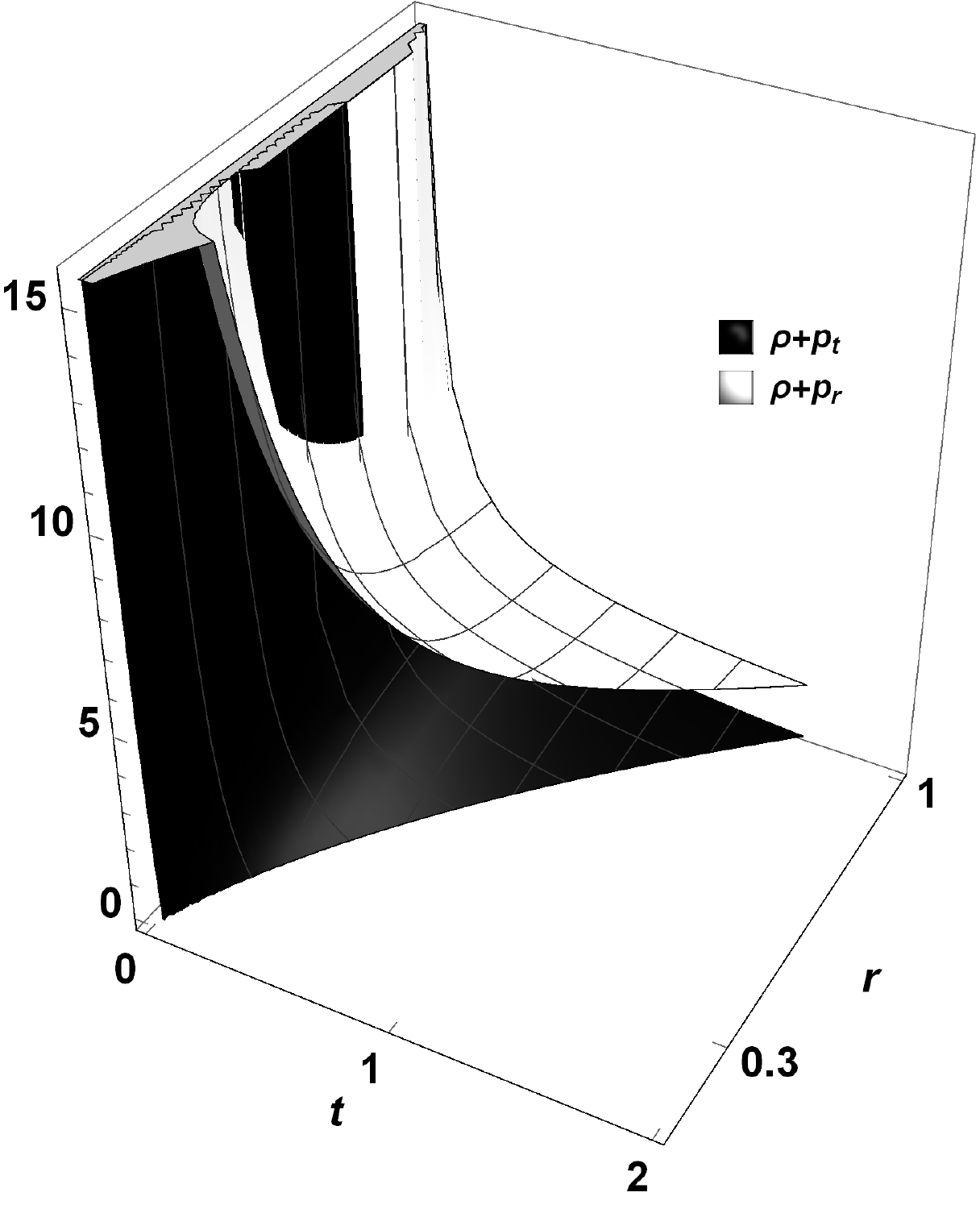}\quad
\includegraphics[width=.25\textwidth]{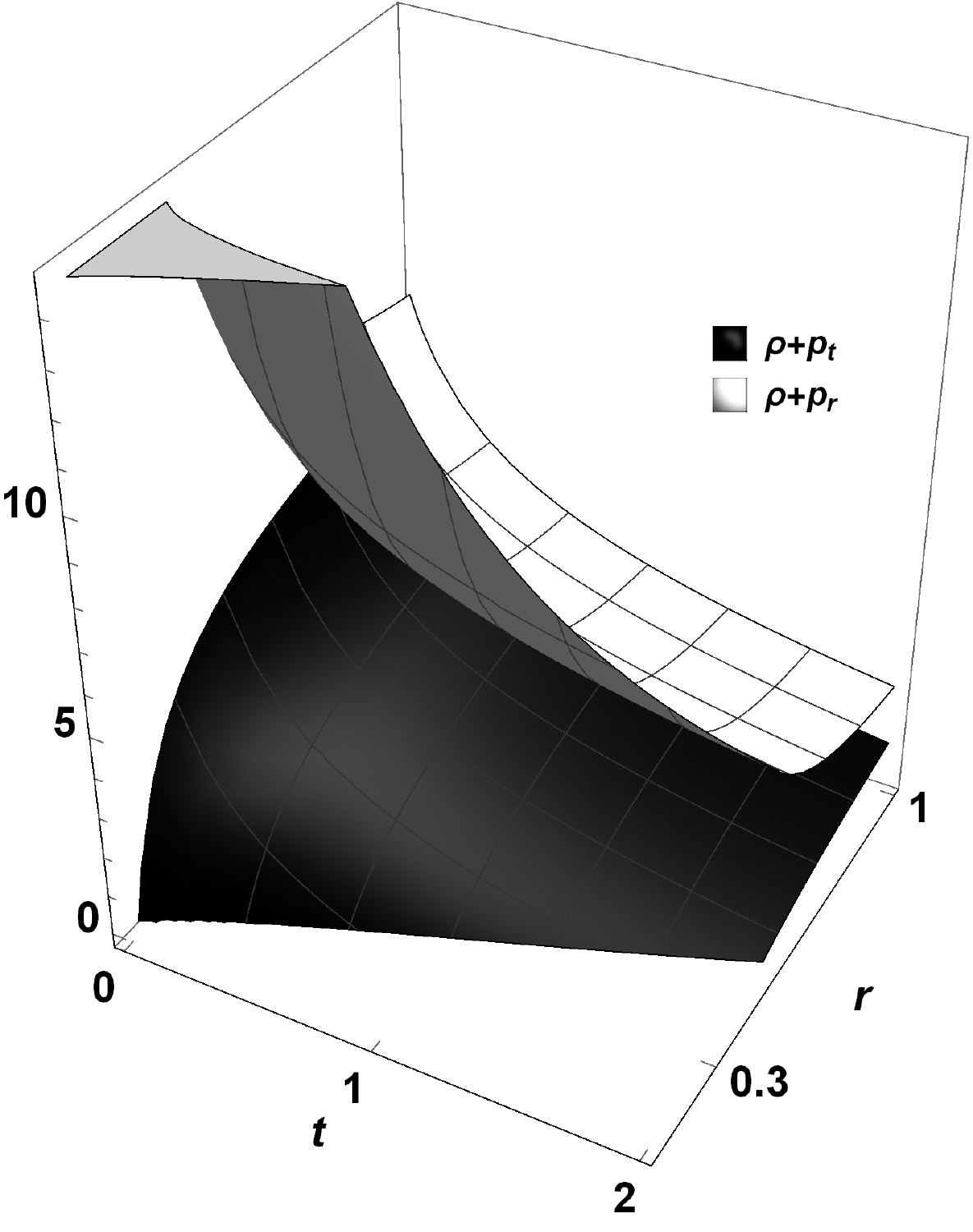}

\caption{The above panels shows the variations in $\rho+ p_{r}$ and $\rho+ p_{t}$ for three different $a(t)$ corresponding to $g^{2}(r)=\left(\frac{1}{1+r-r_{0}}\right)$ subject to the conditions for which NEC is satisfied.}
\end{figure}

\section{Conclusion}

In the above work we have described a general method of obtaining solutions for evolving wormhole geometries. We have further discussed different wormhole solutions and described in details their matter tensor dependence on the temporal factor. We have explicitly shown for the first time, that time dependent evolving wormhole geometries cannot have radially dependent time-time component of the metric tensor. Accordingly we have obtained wormhole solutions with only time dependent 0-0 component metric tensor. 

Solutions similar to those obtained in section 2.1.1 have been discussed in \cite{rom} and \cite{kar}. In \cite{rom}, Roman, used the scale factor $e^{\xi t}$ to describe an evolving wormhole geometry and considered the effects of inflation on such wormholes. In \cite{kar} similar solutions were discussed in the context for wormholes existing with successfully satisfying weak energy conditions (WEC), while in \cite{rom} such wormholes violated WEC. In our paper the various classes of solutions obtained embodies and generalises the solutions of both \cite{rom} and \cite{kar}, in the sense that our wormhole examples can exist both with and without violating the NEC. Such existence is however for some finite arbitrary amount of time depending on several parameter restrictions as shown in the article.  The remaining two solutions discussed in section 2.1.2 are in fact general classes of some of the solutions discussed in \cite{ks}. We reiterate that all the solutions discussed in \cite{ks} can be obtained by the general method suggested in this work. 

Thus we have not only provided a general method of obtaining solutions for evolving wormholes, we have also shown such wormholes can exist with throat matter satisfying the energy conditions. Further all our solutions are cosmological in nature and can exist in different cosmological scenarios. We have also successfully interpreted the constant $\beta$ as the physically significant parameter indicating the curvature of the universe. Significantly it appears due to the relation between the two stress tensors. Finally we have validated all our claims using numerical evaluations through graphical representations. 
\section{Acknowledgments}

SB and TB thanks {\it Inter University Centre for Astronomy and Astrophysics(IUCAA)} for their research facility and hospitality, where this work was initiated. SB acknowledges UGC's Faculty Recharge Programme and Department of Science and Technology, SERB, India for financial help through ECR project (File No. ECR/2017/000569).

\section{Data Availability}
Data sharing is not applicable to this article as no new data were created or analyzed in this study.

\end{document}